# Experimental Study of Forced Synchronization and Cross-coupling in a Liquid-Fuelled Gas Turbine Combustor at Elevated Pressure

Mitchell Passarelli[a,*], Askar Kazbekov[a], Victor Salazar[b], Krishna Venkatesan[b], Adam M. Steinberg[a]

[a]School of Aerospace Engineering, Georgia Institute of Technology, 270 Ferst Dr, Atlanta, GA 30332, USA
[b]GE Research, 1 Research Cir, Niskayuna, NY 12309, USA

______________________________________________________________________

**Abstract**

The effects of external forcing on a turbulent, liquid-fuelled, swirl-stabilized gas turbine combustor operating at a pressure of approximately 1 MPa are explored experimentally. In particular, the dynamics and coupling between the hydrodynamics, heat release rate and acoustics are compared for various forcing amplitudes at a fixed forcing frequency $f_f$. The hydrodynamics were characterized via laser Mie scattering from droplets in the fuel spray, while the heat release rate was qualitatively measured using chemiluminescence (CL) emissions in the 312 ± 12.5 nm wavelength range, both at 10 kHz. The dynamics at the frequencies of interest were extracted using spectral proper orthogonal decomposition (SPOD). In the unforced case, the spray and CL oscillations exhibited similar dynamics, dominated by oscillations at frequency $f_0$, whereas the pressure fluctuations were predominantly at $f_P$. As the forcing amplitude was increased from zero, the spray and CL exhibited changes in their power spectra characteristic of the suppression route to synchronization. The pressure fluctuations, however, were observed to follow the phase-locking route to synchronization. In contrast with expectations from synchronization theory, the amplitude of the pressure fluctuations increased significantly not only after lock-on, but also as the frequency detuning with $f_f$ decreased. It is shown that this increase in amplitude is not due to intermittency in the frequency of the pressure oscillations. The simultaneous occurrence of phase-locking and suppression illustrates the rich variety of dynamics that can occur in practical combustor systems. In addition, the amplification of the pressure oscillations based on the frequency detuning with the forcing suggests that classical reasoning based on the Rayleigh Index may not be sufficient to understand the high amplitude behaviour of multimodal systems.

______________________________________________________________________

*Corresponding author.

# 1. Introduction

This paper articulates the routes to synchronization between forced and self-excited oscillations in the spray, flame and pressure for a liquid-fuelled gas turbine model combustor at elevated pressure. Gas turbine combustors exhibit oscillatory dynamics in a variety of physical phenomena at a variety of frequencies [1, 2]. For example, global hydrodynamic instabilities result in coherent velocity oscillations, the natural frequencies of which can be predicted through linear stability analysis. Meanwhile, pressure oscillation frequencies depend on the acoustic modes of the combustor and can potentially lead to coupled oscillations in the velocity, fuel flow, etc. Both of these phenomena can lead to associated oscillations in the heat release rate at their respective frequencies via various complex coupling pathways [3-9].

As first discussed by Rayleigh [9], self-excited thermoacoustic instabilities arise when the pressure and heat release rate oscillate sufficiently in-phase with one another. Existing studies [10-13] tend to consider only dynamics at the same frequency. With traditional mathematical formulations of the Rayleigh Criterion, such as the Rayleigh Index for example, a helical shear layer instability mode at a particular frequency would not amplify pressure and heat release dynamics at a different (even slightly detuned frequency) due to the orthogonality of sine waves.

However, the dynamics of nonlinear oscillatory systems allows various routes through which different phenomena at different frequencies can interact and ultimately synchronize. This leads to the possibility of, for example, hydrodynamic modes at a given natural frequency synchronizing with and exciting acoustic modes at a different frequency. On the other hand, it also allows for the possibility of active control of thermoacoustic instabilities through forcing, even if the forcing frequency is not perfectly tuned to the natural instability frequency [14, 15]. Indeed, it is worth noting that synchronization can occur in frequency bands around any integer ratio of the natural frequencies of the phenomena of interest [16]. The rich dynamics of such self-oscillating systems require careful analysis and experimentation to understand their response and instability mechanisms.

The dynamics of forced self-oscillators are similar to those of a pair of mutually coupled self-oscillators as described in Refs. [5, 16]. The main difference is that the forcing is typically externally driven, and thus does not respond to changes in the system to which it is applied, *i.e.*, there is a one-way coupling that allows only the forcing to act on the self-oscillator. The unidirectionality of the coupling differentiates the forced response from the dynamics resulting from mutual coupling of multiple self-excited modes.

Consider an oscillator with natural frequency $f_n$ subjected to an external forcing at frequency $f_f$ and with amplitude $F$ such that $f_n/f_f \approx m/n$ for integer values of $m$ and $n$. The generalized frequency detuning $\Delta f^n = mf_f - nf_n$ expresses the difference between the frequencies with respect to the nearest synchronization tongue. In systems where the natural frequencies cannot be accurately determined, the observed frequency ($f_o$) can approximate the frequency detuning, *i.e.*, $\Delta f^n \approx \Delta f = mf_f - nf_o$. The interactions between the forcing and the natural oscillations can cause $f_o$ to differ from $f_n$, thus requiring additional information to identify the synchronization state of the system for forced oscillations (discussed further in Section 3).

In general, the behaviour of the oscillations is determined by both $\Delta f^n$ and the forcing amplitude, $F$. Here, the forcing amplitude is analogous to the dissipative and reactive coupling strengths in a mutually coupled system [16]. Since the coupling with the forcing is one-way, the possible states differ from those observed in mutually coupled systems. Except for infinitesimal forcing amplitudes, which yield a linear superposition of the unforced and forced solutions, the presence of forcing always interacts with the observed frequency of the system [16].

That is, for a given pair $(\Delta f^n, F)$, there are three possible system states: 1.) unsynchronized and coupled oscillations; 2.) phase synchronized or phase-locked oscillations and; 3.) suppressed dynamics. The parameter space over which synchronization occurs for harmonic ratio $m:n$ is roughly V-shaped, as illustrated schematically in Fig. 1.

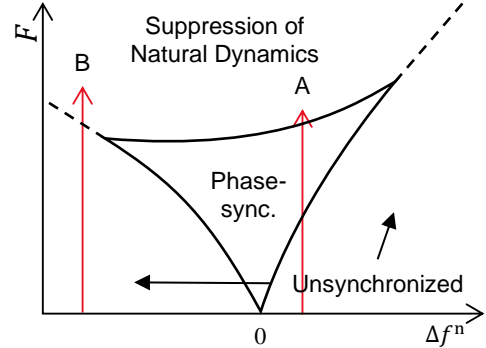

Fig. 1. Schematic representation of $m:n$ forced synchronization region. Arrow A denotes the phase-locking route; B denotes the suppression route.

The red arrows in Fig. 1 illustrate the two possible routes to synchronization for a given $\Delta f^n$. Arrow A lies along the phase-locking route, which occurs for relatively small values of $\Delta f^n$. Starting from zero forcing amplitude, as the forcing amplitude increases the observed frequency $f_o$ is pulled towards $f_f$ until they coincide. The amplitude of the 'natural' dynamics does not change.

Arrow B in Fig. 1 denotes the suppression of natural dynamics route and occurs for values of $\Delta f^n$ larger than those for the phase-locking route. Increasing the forcing amplitude along this route has little effect on the frequency of the 'natural' dynamics. Instead, the amplitude of the natural oscillations decreases to zero as the response at the forcing frequency increases. Synchronization is achieved once the amplitude of the natural dynamics vanishes.

Classical intuition based on mathematical formulations of the Rayleigh Criterion (such as the Rayleigh Index) may be insufficient to capture cross-frequency coupling related to synchronization phenomena. The Rayleigh Index posits that energy transfer between the acoustic and heat release fields in a combustor only occurs when the pressure and heat release rate oscillations are at the same frequency. The interactions involved with the suppression of natural dynamics, for example, involve oscillations at distinct frequencies. As such, these types of interactions may limit the ability of the Rayleigh Index to determine and understand whether a multimodal combustor system will exhibit high amplitude dynamics.

Existing experimental studies [6, 8, 17-21] have explored the effects of forcing on premixed combustors operating with gaseous fuels and at atmospheric pressure. Their research shows that the forcing can interact with natural acoustic oscillations at nearby frequencies, inducing changes in frequency and/or amplitude. In addition, the use of synchronization and bifurcation theory provides a framework for understanding and classifying these dynamical changes. The work of Guan et al. [20], for example, illustrates the phase-locking route in response to forcing, while Murugesan et al. [21] explore the suppression of natural dynamics.

Other groups have investigated how forcing can impact natural hydrodynamic instabilities. Lückoff and Oberleithner [22] experimentally demonstrate how applied acoustic forcing interacts with a precessing vortex core. Their results show that the acoustic forcing can suppress the oscillations associated with hydrodynamic instabilities, having a significant effect on the observed flame and flow structure. The specific mechanisms of such interactions are also explored by Terhaar et al. [23].

Forced dynamics and synchronization phenomena are well-researched and understood across many systems, *e.g.*, Refs. [14, 24-26]. However, there are a limited number of experimental studies [14, 27-29] that investigate such phenomena in liquid-fuelled systems and fewer that operate at pressures and power-densities typical in practical combustors.

Exploring forced synchronization phenomena at flight-relevant conditions for liquid-fuelled gas turbine combustors is primarily motivated by this lack of data. As numerical simulations increase in fidelity, it is important to ensure that they can accurately resolve or model all phenomena. The work of Chen *et al.* [30] illustrates how computational fluid dynamics (CFD) simulations must be set up correctly to properly resolve synchronization phenomena. These simulations, however, are of a simplified thermoacoustic engine, not a gas turbine combustor system. Lo Schiavo *et al.* [29] remark that synchronization can also significantly impact the simulation and modelling of liquid spray characteristics. Furthermore, the works of Kopasakis *et al.* [14], Guan *et al.* [17], Lückoff and Oberleithner [22] and Chen *et al.* [30] explore potential applications for active control methods. For any active control strategy to work in practice, the underlying physics of the system to be controlled must be understood. As Sujith and Unni [15] note, any control method must be designed to account for the myriad of interactions that can occur with the natural dynamics.

To this end, this paper presents forced synchronization experiments of a liquid-fuelled model aeronautical gas turbine combustor operating at flight-relevant conditions. Through a sweep of the forcing amplitude at a fixed forcing frequency, both the phase-locking and suppression routes to synchronization are observed. In addition, the collected data indicates that the natural mutual coupling between the pressure, flame and hydrodynamics can interfere with phase-locking. As a result, the observed forced response is more complex than expected.

## 2. Experimental setup

The experiments were performed at GE Research on an optically accessible model gas turbine combustor, a schematic of which is shown in Fig. 2. The combustor operates at fuel-rich conditions using liquid Jet A fuel supplied via a swirl nozzle. Experiments were performed at air preheat temperatures of approximately 500 K and pressures above 1 MPa with a thermal power of approximately 0.5 MW. The combustor was allowed to reach a statistically stationary operating point prior to data acquisition, which was verified using thermocouple, static pressure and mass flow rate measurements.

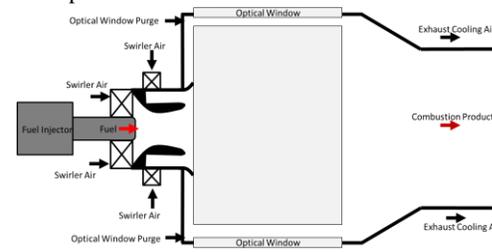

Fig. 2. Schematic of test section in planform view.

For the experiments presented in this paper, a siren was connected to the air feed lines to apply forcing at specific frequencies and amplitudes. Data were recorded for an unforced case and three forced cases at the same forcing frequency (denoted by $f_f$) but with varying forcing amplitudes (denoted by $F$).

Measurements were made of the flame, fuel spray and pressure dynamics using high-speed chemiluminescence (CL) imaging, Mie scattering and pressure transducers, respectively. CL imaging of emissions in the wavelength range $312 \pm 12.5$ nm was recorded at 10 kHz using a high-speed camera (Photron SA5) equipped with a bandpass filter, commercial objective lens (Nikkor 105 mm UV, f/# = 16) and lens-coupled image intensifier (Invisible Vision UVi 2550B-10-S20, gate time of 40 μs). Simultaneous wide-spectrum measurements of the spatially- and temporally-integrated combustor emissions made using a portable spectrometer indicated that the signal in the collected wavelength range corresponded to broadband CL emissions from $CO_2$* and other large molecules; the signal cannot be attributed solely to OH*. As such, the CL images only provide a qualitative, line-of-sight integrated indicator of the flame fluctuations [3].

Fuel dynamics were characterized based on laser Mie scattering from the droplets. The flow was illuminated using the second harmonic output of a Nd:YAG laser (Quantronix Hawk-Duo), with pulses repeating at 10 kHz. Three cylindrical lenses formed the beam into a sheet with a thickness of approximately 2 mm at the beam-waist in the test section. The imaging system consisted of two high-speed cameras (Phantom v2640) equipped with 532 ± 2.5 nm bandpass filters, objective lenses (Tamron 180 mm, f/# = 11) and Scheimpflug adapters (LaVision). We note that the two-camera system was employed to enable stereoscopic velocimetry measurements, which are not reported here. Hence, the data in this paper is taken from a single camera.

Dynamic pressure transducers (PCB) measured the pressure fluctuations within the combustor at two axial locations. The transducers were mounted to the ends of waveguides in a semi-infinite loop configuration to isolate them from the temperatures in the combustor. The analysis presented in this paper is not affected by time lags or attenuation induced by the waveguides. However, there is a slight uncertainty in the reported frequencies. Pressure signals and timing signals from the camera were recorded using a data acquisition system (National Instruments USB-6361) at a rate of 100 kHz, which was sufficiently high to prevent any aliasing for the oscillation frequencies discussed below. The two pressure measurements were qualitatively identical; thus, the more upstream transducer was used for the analysis presented here.

## 3. Analysis techniques

As mentioned in Section 1, $\Delta f$ alone is not sufficient to determine the synchronization state of a forced self-oscillator. The changes in the power spectra as the amplitude $F$ is varied for a fixed $f_f$ can also provide a qualitative indication of the route of synchronization and the current synchronization state of the system. For more quantitative analysis, one can use metrics derived from the generalized phase difference $\Delta \phi = n\phi_f - m\phi_o$, where $m$ and $n$ are from the harmonic ratio of the frequencies of oscillation as in Section 1, and $\phi_f$ and $\phi_o$ denote the oscillation phase of the forcing and observed dynamics.

It is possible to inspect plots of the temporal variation of $\Delta \phi$ to identify the synchronization state of the system [6, 18]. The effects of noise and nonstationarity, however, can limit the usefulness of such plots. The phase-locking value (PLV) is a popular statistical alternative to quantify the degree of phase-locking [31, 32]. The PLV over a given time interval $T$ is defined as $PLV = 1/T \left| \sum_{t=1}^{T} e^{i\Delta\phi(t)} \right|$. The PLV is a quantitative measure of the nonuniformity of the distribution of $\Delta \phi$ which indicates synchronization of coupled oscillators [31].

The main difficulty with the PLV is determining the threshold between phase-locked and unsynchronized dynamics. For the analysis presented below, the system is considered phase-locked when $PLV \gtrsim 0.9$. In addition. the PLV is implemented using a sliding window to capture behaviours that are intermittent in time.

Given the multimodal and nonstationary nature of the data collected, it is important to be able to extract the dynamics corresponding to frequencies of interest. To do so, the spectral proper orthogonal decomposition (SPOD) of Sieber *et al.* [33] was selected. The SPOD algorithm is nearly identical to that of classical POD, with the additional step of lowpass filtering along the diagonals of the correlation matrix prior to the eigenvalue decomposition. This filtering operation acts like a spectral bandpass filter on each mode, redistributing the energy outside the passband to other modes. Thus, SPOD avoids the mode blending exhibited by POD without loss of information like in bandpass filtering or phase-averaging [33]. As a result, the SPOD modes oscillate within distinct frequency ranges and are more robust to nonstationarity and noise than modes from dynamic mode decomposition (DMD).

Using the mode pairing algorithm described by Sieber *et al.* [33], the temporal coefficients of each mode in a pair can be used to construct the analytical signal for that pair. In this manner, the phase $\phi$ of the oscillations can be computed for a specific mode, analogous to using the Hilbert Transform [34, 35]. Furthermore, the cross-comparison of these modes in isolation is sufficient for exploring their coupling as discussed in Ref. [16]. In other words, without loss of generality, it is possible to analyze only the pairwise interactions between two frequencies that have been extracted from multimodal dynamics [16].

## 4. Results & discussion

To begin, the SPOD spectra of the CL, spray and pressure fluctuations were computed for the unforced case to determine the characteristics of the natural modes in the combustor. These SPOD spectra are shown in Fig. 3, wherein important or significant coherent structures in the data are denoted by large, bright coloured dots. Note that the frequency in these plots is normalized by $f_f$, which is slightly greater than $f_0$. Figure 3 illustrates two interesting points. First, the CL and spray oscillate at the same frequencies (denoted $f_0$ and $2f_0$) and have the same helical mode shapes; the mode shapes are articulated in Fig. 4.

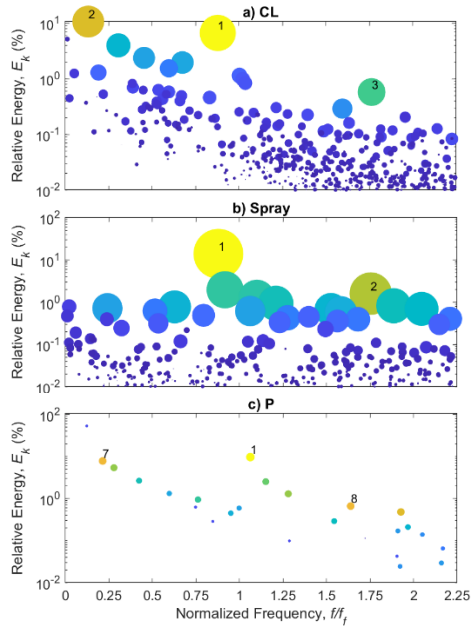

Fig. 3. SPOD spectra of unforced a) CL, b) spray and c) pressure. Numbers denote spectral coherence, where "1" is the most coherent.

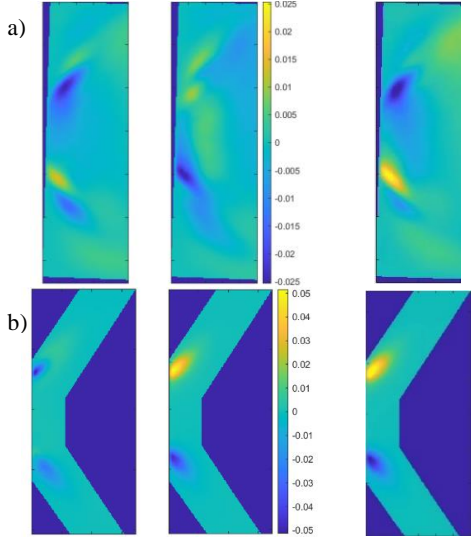

Fig. 4. SPOD mode shapes of unforced dynamics at $f_0$ for a) CL and b) spray. Third column is a snapshot of the reconstructed dynamics.

As will be discussed below, the spray and CL exhibit the same mode shapes and frequencies, even in the forced cases; the CL dynamics are closely linked to that of the fuel spray. Hence, only the spray data are presented for the subsequent analysis and comparisons. Second, Fig. 3 reveals that the pressure oscillates at frequencies (predominantly $f_P \approx 1.2 f_0 \approx 1.05 f_f$) distinct from those in the spray and CL.

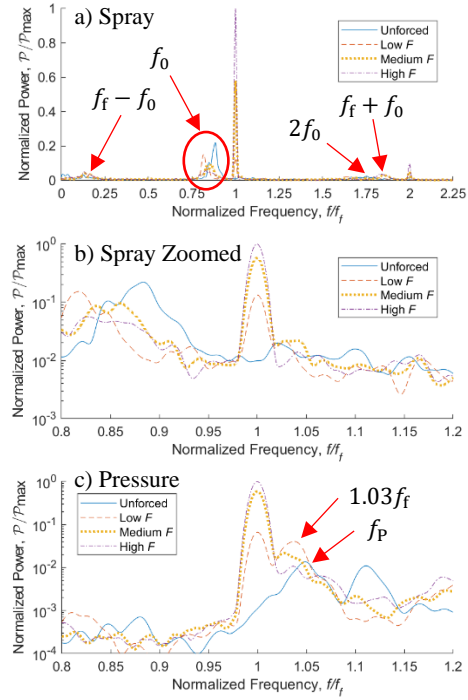

Fig. 5. Normalized power spectra for a) spray, b) spray zoomed-in and c) pressure. Spectra in b) and c) are zoomed-in and have logarithmic y-axes to focus on changes near $f_f$. Powers are normalized by maximum across all test cases. Frequencies of interest are labelled.

The effects of the forcing at various amplitudes are shown in the power spectra of Fig. 5. The forcing frequency $f_f \approx 1.12 f_0 \approx 0.95 f_P$ was held constant for each case, varying only the amplitude of the forcing. The power spectra were computed using Welch's method of overlapping windowed segments, using Hamming windows sized to optimize the balance between spectral resolution and noise.

Consider first the spray oscillations in Fig. 5a-b. At the lowest forcing amplitude, the power spectrum of the spray includes an additional mode at $f_f$. As the forcing amplitude increased (cf. Fig. 5b), the amplitude of the forced oscillations increases, while the amplitude of the oscillations at $f_0$ decreases. Since the frequency of the spray oscillations near $f_0$ is not drawn towards $f_f$ as the forcing amplitude increases, the observed behaviour indicates that the system is thus progressing along the suppression route (Arrow B in Fig. 1). Given that the natural oscillations have a nonzero amplitude even at the highest forcing amplitude, the natural dynamics are not suppressed in any of the cases. The appearance of sum and difference frequencies in the forced cases is due to the inharmonicity of the oscillations [16]. Note that the changes in frequency of the oscillations near $f_0$ are due to the changes in combustor inlet temperature across the test cases.

In contrast to the CL and spray, the pressure does not exhibit suppression-like behaviour. As the forcing amplitude increases, the frequency of the most dominant mode is drawn towards $f_f$ until it matches (Fig. 5c). For the lowest forcing amplitude, the pressure spectrum contains peaks both at the forcing frequency $f_f$ and between the forcing frequency and unforced pressure frequency, *i.e.*, the frequency-pulled natural dynamics are at $1.03f_f \approx 0.97f_P$. It is worth noting that the oscillations at $f_f$ are rather incoherent based on the SPOD analysis. The mode corresponding to the frequency-pulled natural dynamics is the most coherent in the pressure at the low forcing amplitude. Increasing the forcing amplitude causes the pressure to oscillate at $f_f$.

This frequency-pulling is characteristic of the phase-locking route to synchronization (Arrow A in Fig. 1). Indeed, the frequency detuning between the unforced pressure and the forcing ($\Delta f_{Pf} = f_P - f_f \approx 0.05f_f$) is less than that between the unforced CL or spray and the forcing ($\Delta f_{fs} = f_f - f_0 \approx 0.12f_f$). These frequency detuning values are also consistent with those corresponding to phase-locking and suppression reported in other works [6]. To verify the phase-locked state of the pressure for higher amplitudes of forcing (when the pressure oscillated at $f_f$), the PLV is plotted in Fig. 6. Fig. 6a contains the PLV between the pressure and CL oscillations at $f_f$, checking for phase-locked thermoacoustic oscillations, while Fig. 6b is between the pressure and the spray, revealing any hydrodynamic coupling. At low forcing amplitudes, the pressure appears to lock on intermittently, while at medium and high forcing amplitudes, the pressure is completely phase-locked to the forcing in both the CL and spray.

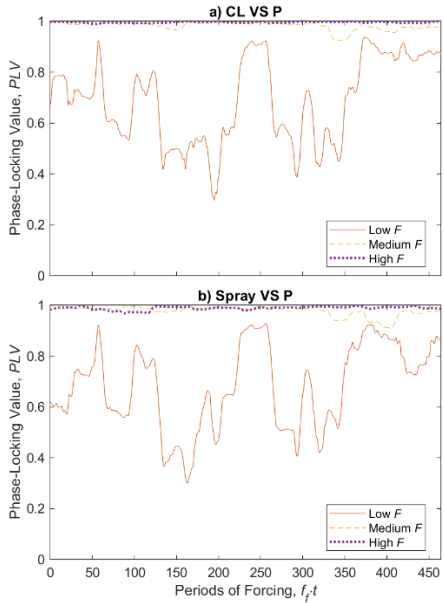

Fig. 6. Phase-locking value between $f_P$ pressure and $f_f$ a) CL and b) spray SPOD modes for each forcing amplitude.

However, there is an interesting difference between the response of the pressure to the forcing (as shown in Fig. 5c) and the nominal theoretical description of phase-locking; whereas the amplitude of the natural dynamics nominally should remain relatively constant along the phase-locking route to synchronization, Fig. 5c shows a nearly threefold increase in amplitude for the lowest forcing case compared to the natural dynamics. Given that the forcing pulls the frequency of the natural pressure oscillations towards $f_f$, it was first hypothesized that the increase in amplitude was due to an intermittent Rayleigh Index effect. As $F$ increased, $\Delta f$ between the natural pressure and forced CL oscillations decreased, which would increase the amount of time the pressure and CL oscillations were in-phase. To evaluate this hypothesis, the time evolution of the pressure power spectra for the low forcing amplitude case was inspected using the spectrogram in Fig. 7. Figure 7 indicates that the time intervals where the pressure frequency drifts towards $f_f$ coincide with times in Fig. 6 when the PLV is low, suggesting the oscillations are not phase-locked and thus not interacting strongly in a Rayleigh Index sense. This suggests that the forcing enables an additional energy transfer pathway between different frequencies.

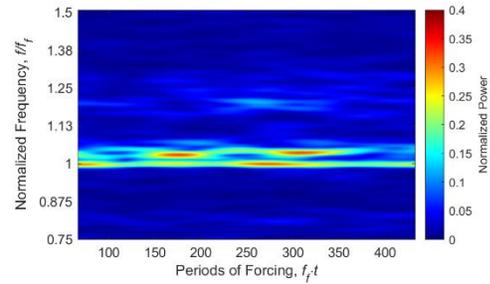

Fig. 7. Normalized spectrogram of pressure subjected to low amplitude forcing. Truncated to region of interest.

Another interesting difference between theoretical expectations of the phase-locking response and the observations in Fig. 5c is the presence of oscillations near $f_P$ for all forcing amplitudes. For a noise-free, single-mode self-oscillator, synchronization theory predicts that the phase-locked forced response should not have any significant frequency content at the natural frequency [16]. The dynamics examined here and reported in Fig. 5, however, are taken from a forced, noisy, nonstationary and multimodal combustor system, consisting of several oscillating subsystems (*e.g.*, pressure, flame and fuel spray dynamics) with mutual coupling between them. Thus, there are several additional factors that could give rise to this behaviour. These factors are explored below as possible explanations for why the phase-locked dynamics in Fig. 5c retain some spectral content near $f_P$.

The existence of oscillatory dynamics at other frequencies, along with the mutual coupling between the spray, CL and pressure, increases the complexity of the synchronization phenomena. It is likely that the natural dynamics arise, at least in part, due to synchronization phenomena between the various frequencies in the combustor. As noted by Guan *et al.* [20], the mutual coupling between these modes must first be overcome by the forcing before individual forced synchronization can occur. In addition, the resistance of the mutual coupling can also yield several additional dynamical states that are not observable for a forced single-mode self-oscillator [20]. Given that the natural frequency of the spray oscillations, $f_0$, (cf. Fig. 5a-b) changes across the test cases, the effects of the mutual interactions between $f_0$ and $f_P$ will likely differ as well. In their experiments, Guan *et al.* [20] also observe a state where the natural dynamics can desynchronize from the forcing for high enough forcing amplitudes.

The presence of high energy turbulent fluctuations and noise further complicates the response compared to basic forced synchronization. Specifically, intense noise both interferes with synchronization and also induces an ordered motion near the natural frequency via coherence resonance [16]. Coherence resonance manifests as a spectrally broad peak, relative to the bandwidth of the forcing, near the natural frequency of the oscillator and the height of the peak depends on the forcing amplitude.

## 5. Conclusion

The effects of forcing on the natural dynamics of a liquid-fuelled gas turbine combustor operating at flight-relevant conditions were recorded experimentally. As expected, increasing the forcing amplitude at a constant forcing frequency causes the natural hydrodynamics (indicated via fuel spray Mie scattering measurements), heat release rate oscillations (via broadband CL emissions) and acoustics to synchronize with the forcing. The forcing amplitude required to induce synchronization is observed to depend on the frequency detuning between the forcing and the natural oscillation frequencies.

For these experiments, the fuel spray and CL dynamics exhibited behaviour following the suppression route to synchronization. Even at the highest forcing amplitude, however, the natural dynamics remained unsuppressed. In contrast, the pressure dynamics, whose natural frequency differed from both the CL and spray, synchronized with the forcing via the phase-locking route. Phase-locking was achieved at moderate forcing amplitudes. In addition, the natural dynamics of the pressure locked onto the forced dynamics in the spray and CL. Finally, it was also observed that the phase-locking caused a significant increase in the power of the pressure oscillations, contrary to predictions from synchronization theory.

The observations have several implications with respect to the forced response of combustor dynamics. While the effects of forcing and synchronization are well understood, the complexity of practical gas turbine combustors can yield unexpected responses. In the case of active control, this could yield undesired results.

First, the presence of forcing can enable dynamics at different frequencies to interact and amplify one another. This amplification appears to depend on the degree of phase-locking between different modes. As discussed, the presence of forcing opens an additional energy pathway between the acoustic, hydrodynamic and flame fluctuations at different frequencies. While the Rayleigh Index captures the energy transfer between oscillations at the same frequency it does not work for cross-frequency interactions. It may be possible to extend the Rayleigh Index to higher orders via generalized averaging techniques commonly employed in the approximation of solutions to systems of nonlinear ODEs.

Second, it is important for simulations and models to accurately capture or resolve the plethora of cross-frequency interactions possible in practical combustor systems. Such phenomena are influenced by both oscillations at any frequency and noise interacting with the natural and forced dynamics of a system.

## Acknowledgements

This work was supported by GE. M. Passarelli acknowledges the support of a NSERC PGS-D fellowship.